%% file: main.tex
\documentclass[envcountsame,runningheads]{llncs}
\usepackage{citesort}
\usepackage{amssymb}
\usepackage{amsmath}
\usepackage[nompar,draft]{commenting}

\declareauthor{al}{Anthony}{blue!90!black}
\authorcommand{al}{comment}
\declareauthor{mh}{Matt}{orange!90!black}
\authorcommand{mh}{comment}

\input{comm}

\input{namedenvs}

\let\llncssubparagraph\subparagraph
\let\subparagraph\paragraph
\usepackage[compact]{titlesec}
\let\subparagraph\llncssubparagraph

\titlespacing*{\subsubsection}{0pt}{0.8\baselineskip}{0.8\baselineskip}

\newcommand\shortlong[2]{#2}

\begin{document}
\title{Decidable models of integer-manipulating programs
    with recursive parallelism \shortlong{}{(technical report)}}
\titlerunning{Integer-manipulating programs with recursive parallelism}
\author{Matthew Hague and Anthony Widjaja Lin}
\institute{
    Royal Holloway, University of London, UK
    \and
    Yale-NUS College, Singapore
}

\maketitle

\input{abstract}

\input{intro}

\input{preliminaries}

\input{model}

\input{weak}

\input{senescent}

\input{conclusion}

\paragraph{Acknowledgments}
This work was supported by the Engineering and Physical Sciences Research
Council [EP/K009907/1] and Yale-NUS College Startup Grant.

\shortlong{
    \input{short-refs}
}{
    {
    \scriptsize
    \bibliographystyle{abbrv}
    \bibliography{references}
    }

    \appendix

\input{senescent-appendix}
}
\end{document}

%% file: comm.tex
\usepackage{ifthen}
\usepackage{xspace}

\newcommand{\OMIT}[1]{}

\newcommand{\defn}[1]{\textit{#1}} 

\newcommand{\sem}[1]{[\![#1 ]\!]}

\newcommand{\sGTRS}{}
\def\sGTRS/{sGTRS}

\newcommand{\GTRS}{}
\def\GTRS/{GTRS}

\newcommand{\rbGTRS}{}
\def\rbGTRS/{rbGTRS}

\newcommand{\N}{\ensuremath{\mathbb{N}}}
\newcommand{\Z}{\ensuremath{\mathbb{Z}}}

\newcommand{\vecM}{\ensuremath{\text{\rm \bfseries m}}}
\newcommand{\vecB}{\ensuremath{\text{\rm \bfseries b}}}

\newcommand{\vecX}{\ensuremath{\text{\rm \bfseries x}}}
\newcommand{\vecY}{\ensuremath{\text{\rm \bfseries y}}}
\newcommand{\vecZ}{\ensuremath{\text{\rm \bfseries z}}}

\newcommand{\eword}{\varepsilon}
\newcommand{\lang}{\mathcal{L}}

\newcommand{\brac}[1]{\left(#1\right)}
\newcommand{\tup}[1]{\brac{#1}}
\newcommand{\ap}[2]{{#1}\mathord{\brac{#2}}}
\newcommand{\set}[1]{\left\{#1\right\}}
\newcommand{\setcomp}[2]{\left\{{#1}\ \middle|\ {#2}\right\}}
\newcommand{\compop}{\sim}

\newcommand{\alphabet}{\ensuremath{\Sigma}}
\newcommand{\rank}{\rho}
\newcommand{\tdom}{D}
\newcommand{\tnode}{\eta}
\newcommand{\tree}{t}
\newcommand{\tlab}{\lambda}
\newcommand{\cha}{a}
\newcommand{\chb}{b}

\newcommand{\cvar}{x}
\newcommand{\context}{C}

\newcommand{\trees}[1]{\mathcal{T}_{#1}}
\newcommand{\tap}[2]{\ap{#1}{#2}}
\newcommand{\csub}[2]{{#1}\mathord{\left[{#2}\right]}}

\newcommand{\ta}{\mathcal{T}}
\newcommand\tastates{\mathcal{Q}}
\newcommand\tarules{\Delta}
\newcommand\tafinals{\mathcal{F}}
\newcommand\tarule[3]{\ifthenelse{\equal{#1}{}}{}{\tup{#1}} \xrightarrow{#2} {#3}}
\newcommand\tast{q}
\newcommand\tarun{\pi}
\newcommand\tasingle[1]{\ta_{#1}}

\newcommand\gtrsrun{\pi}
\newcommand{\gtrs}{G}
\newcommand{\controls}{\mathcal{P}}
\newcommand{\rules}{\mathcal{R}}
\newcommand{\control}{p}
\newcommand{\rrule}[5]{\tup{{#1}, {#2}} \xrightarrow{#3} \tup{{#4}, {#5}}}
\newcommand{\rruler}{\tau}
\newcommand{\config}[2]{\tup{{#1}, {#2}}}
\newcommand{\oalphabet}{\ensuremath{\Gamma}}
\newcommand{\osym}{\gamma}
\newcommand{\tran}[1]{\xrightarrow{#1}}
\newcommand{\run}[1]{\xrightarrow{#1}}
\newcommand{\empsym}{\varepsilon}
\newcommand{\nfatrans}{\Delta}

\newcommand{\ctrs}{C}
\newcommand{\ctr}{c}
\newcommand{\con}{\theta}
\newcommand{\contrue}{\top}
\newcommand{\vals}{\nu}
\newcommand{\dvals}{\mu}
\newcommand{\ccons}[1]{\mathrm{Cons}_{#1}}
\newcommand{\ctrrule}[7]{\tup{{#1}, {#2}, {#3}}
                         \xrightarrow{#4}
                         \tup{{#5}, {#6}, {#7}}}
\newcommand{\ctrconfig}[3]{\tup{{#1}, {#2}, {#3}}}
\newcommand{\val}{v}
\newcommand{\conf}{\alpha}
\newcommand{\phase}{A}

\newcommand{\agebound}{r}
\newcommand{\treein}{\tree^{\mathrm{in}}}
\newcommand{\treeout}{\tree^{\mathrm{out}}}

\newcommand{\cm}{\mathcal{M}}
\newcommand{\cmcontrols}{\mathcal{S}}
\newcommand{\cmcontrol}{s}
\newcommand{\cmrules}{\Delta}
\newcommand{\cminc}[1]{\mathrm{inc}_{#1}}
\newcommand{\cmdec}[1]{\mathrm{dec}_{#1}}
\newcommand{\cmzero}[1]{\mathrm{zero}_{#1}}
\newcommand{\cmrule}[3]{{#1} \xrightarrow[#2]{} {#3}}
\newcommand{\cmop}{\mathit{op}}
\newcommand{\cmconfig}[3]{\tup{#1, #2, #3}}
\newcommand{\cmtran}{\longrightarrow}

\newcommand{\eqcontrol}{\control_=}
\newcommand{\fincontrol}{f}
\newcommand{\cmfreeze}[2]{\tup{#1, #2}}
\newcommand{\nodeint}{\bullet}
\newcommand{\nodegen}{\ast}
\newcommand{\nodenull}{\circ}
\newcommand{\nodectr}[1]{\overline{#1}}
\newcommand{\ctrinc}[1]{\ctr^+_{#1}}
\newcommand{\ctrdec}[1]{\ctr^-_{#1}}
\newcommand{\valszero}{\vals_0}
\newcommand{\dvalsinc}[1]{\dvals^+_{#1}}
\newcommand{\dvalsdec}[1]{\dvals^-_{#1}}
\newcommand{\dvalseq}[1]{\dvals^=_{#1}}
\newcommand{\rulesfresh}{\rules_{\mathrm{fresh}}}
\newcommand{\rulesinc}{\rules_{\mathrm{inc}}}
\newcommand{\rulesdec}{\rules_{\mathrm{dec}}}
\newcommand{\ruleszero}{\rules_{\mathrm{zero}}}
\newcommand{\rulesfin}{\rules_{\mathrm{fin}}}
\newcommand{\numleaves}[2]{\ap{\#_{#2}}{#1}}

\newcommand{\problemx}[3]{
\par\noindent\underline{\sc#1}\par\nobreak\vskip.2\baselineskip
\begingroup\clubpenalty10000\widowpenalty10000
\setbox0\hbox{\bf Instance: }\setbox1\hbox{\bf Question: }
\dimen0=\wd0\ifnum\wd1>\dimen0\dimen0=\wd1\fi
\vskip-\parskip\noindent
\hbox to\dimen0{\box0\hfil}\hangindent\dimen0\hangafter1\ignorespaces#2\par
\vskip-\parskip\noindent
\hbox to\dimen0{\box1\hfil}\hangindent\dimen0\hangafter1\ignorespaces#3\par
\endgroup}

\newcommand{\REG}{\ensuremath{\text{\texttt{REG}}}}
\newcommand{\Modes}{\ensuremath{\text{\textbf{Modes}}}}
\newcommand{\regA}{\ensuremath{A}}
\newcommand{\regB}{\ensuremath{B}}

\newcommand{\Parikh}[1]{\ensuremath{\mathcal{P}(#1)}}

\newcommand{\Dom}{\ensuremath{\text{\texttt{Dom}}}}
\newcommand{\Init}{\ensuremath{\text{\texttt{Init}}}}
\newcommand{\GoodSeq}{\ensuremath{\text{\texttt{GoodSeq}}}}
\newcommand{\Respect}{\ensuremath{\text{\texttt{Respect}}}}
\newcommand{\EndVal}{\ensuremath{\text{\texttt{EndVal}}}}
\newcommand{\EndCounter}[2]{\ensuremath{\text{\texttt{EndCounter}}^{#1}_{#2}}}
\newcommand{\StartCounter}[2]{\ensuremath{\text{\texttt{StartCounter}}^{#1}_{#2}}}

\newcommand{\regVar}[2]{\ensuremath{reg^{#1}_{#2}}}
\newcommand{\revVar}[2]{\ensuremath{rev^{#1}_{#2}}}
\newcommand{\arrVar}[2]{\ensuremath{arr^{#1}_{#2}}}

\newcommand{\Implies}{\ensuremath{\Rightarrow}}

%% file: namedenvs.tex
\newenvironment{namedtheorem}[2]{%
    \expandafter\gdef\csname reftheorem#1\endcsname{%
        Theorem~\ref{#1} (#2)%
    }%
    \begin{theorem}[{#2}] \label{#1}%
}{%
    \end{theorem}%
}
\newcommand\reftheorem[1]{\expandafter\csname reftheorem#1\endcsname}

\newcommand\reflemma[1]{\expandafter\csname reflemma#1\endcsname}

\newenvironment{namedproperty}[2]{%
    \expandafter\gdef\csname refproperty#1\endcsname{%
        Property~\ref{#1} (#2)%
    }%
    \begin{property}[{#2}] \label{#1}%
}{%
    \end{property}%
}
\newcommand\refproperty[1]{\expandafter\csname refproperty#1\endcsname}

\newenvironment{nameddefinition}[2]{%
    \expandafter\gdef\csname refdefinition#1\endcsname{%
        Definition~\ref{#1} (#2)%
    }%
    \begin{definition}[{#2}] \label{#1}%
}{%
    \end{definition}%
}
\newcommand\refdefinition[1]{\expandafter\csname refdefinition#1\endcsname}

%% file: abstract.tex
\begin{abstract}
We study safety verification for 
multithreaded programs with
recursive parallelism (i.e. unbounded thread creation and recursion)
as well as unbounded integer variables.
Since the threads in each program configuration are structured in a
    hierarchical fashion, 
    our model is state-extended ground-tree rewrite systems
    equipped with
    shared unbounded integer counters that
    can be incremented, decremented, and compared against an integer constant.
    Since the model is Turing-complete, we propose a decidable
    underapproximation.
    First, using a restriction similar to context-bounding,
    we underapproximate the global control by a weak global control
    (i.e. DAGs possibly with self-loops), thereby limiting the
    number of synchronisations between different threads.
    Second, we bound the number of reversals between non-decrementing and
    non-incrementing modes of the counters.
    Under this restriction, we show that reachability becomes NP-complete.
    In fact, it is poly-time reducible to satisfaction over existential
Presburger formulas, which allows one to tap into highly optimised SMT solvers.
Our decidable approximation strictly generalises known decidable models
    including (i) weakly-synchronised ground-tree rewrite systems, and
    (ii) synchronisation/reversal-bounded concurrent pushdown systems systems
    with counters. Finally, we show that, when equipped with reversal-bounded
    counters, relaxing the weak control restriction by the notion of senescence
    results in undecidability.

    \OMIT{
Pushdown systems (PDS) naturally model sequential recursive programs.  Numeric
data types also often arise in real-world programs.  We study the extension of
PDS with unbounded counters, which naturally model numeric data types.  Although
this extension is Turing-powerful, reachability is known to be decidable when
the number of reversals between incrementing and decrementing modes is bounded.
In this paper, we (1) pinpoint the decidability/complexity of reachability and
linear/branching time model checking over PDS with reversal-bounded counters
(PCo), and (2) experimentally demonstrate the effectiveness of our approach in
analysing software.  We show reachability over PCo is $\NP$-complete, while LTL
is $\coNEXP$-complete ($\coNP$-complete for fixed formulas). In contrast, we
prove that EF-logic over PCo is undecidable. Our $\NP$ upper bounds are by a
direct poly-time reduction to satisfaction over existential Presburger formulas,
allowing us to tap into highly optimized solvers like Z3.  Although
reversal-bounded analysis is incomplete for PDS with unbounded counters in
general, our experiments suggest that some intricate bugs (e.g. from Linux
device drivers) can be discovered with a small number of reversals.  We also
pinpoint the decidability/complexity of various extensions of PCo, e.g., with
discrete clocks.
}
\end{abstract}

%% file: intro.tex
\section{Introduction}

Verification of multithreaded programs is well-known to be a challenging
problem. One approach that has proven effective in addressing the problem is to
bound the number of context switches \cite{QR05,Q08}. [Recall that a
\defn{context switch} occurs when the CPU switches from executing one thread
to executing a different thread.] When the number of context switches is
fixed, one may adopt pushdown systems as a model of a single thread
and show that reachability for the concurrent extension of the
abstraction (i.e. multi-pushdown systems) is NP-complete \cite{QR05}.
This result has paved the way for an efficient use of highly optimised
SMT solvers in verifying concurrent programs (e.g. see \cite{HL12,AAC13,EGP14}).
Note that without bounding the number of context switches the model
is undecidable \cite{R00}.

In the past decade the work of Qadeer and Rehof \cite{QR05} has spawned a lot of
research in underapproximation techniques for verifying multithreaded programs,
e.g., see
\cite{HL12,AAC13,EGP14,SES08,ABQ11,LTKR08,MQ07,L12,H14,lTMP07,ABH08,AGN14,ANS14,GMM12,MP11,lTNP14,lTMP07,CHL13}
among many others.
Other than unbounded recursions, some of these results simultaneously address
other sources of infinity,
e.g., unbounded thread creation \cite{L12,H14,ABQ11}, unbounded integer
variables \cite{HL12}, and unbounded FIFO queues \cite{AAC13,AGN14}.

\paragraph{Contributions.}
In this paper we generalise existing underapproximation techniques
\cite{HL11,L12} so
as to handle both shared unbounded integer variables and recursive parallelism
(unbounded thread creation and unbounded recursions). The paper also provides
a cleaner proof of the result in \cite{HL12}: an NP upper bound for
synchronisation/reversal-bounded reachability analysis of concurrent pushdown
systems with counters. We describe the details below.

We adopt state-extended
ground-tree rewrite systems (\sGTRS/) \cite{L12}
as a model for multithreaded programs with
recursive parallelism (e.g. programming constructs including
 \texttt{fork}/\texttt{join}, \texttt{parbegin}/\texttt{parend},
and \texttt{Parallel.For}). Ground-tree rewrite systems (\GTRS/) are known
(see \cite{GL11}) to strictly subsume other well-known sequential and
concurrent models like pushdown systems \cite{BEM97}, PA-processes \cite{EP00},
and PAD-processes \cite{mayr-thesis}, which are known
to be suitable for analysing concurrent programs. [One may think of \GTRS/ as
an extension of PA and PAD processes with return values to parent threads
\cite{GL11}.] We then equip \sGTRS/ with unbounded integer counters that can
be incremented, decremented, and compared against an integer constant.

Since our model is Turing-powerful, we provide an underapproximation of the
model for which safety verification becomes decidable.
First, we underapproximate the global control by a weak global control
\cite{KRS04,L12}
(i.e. DAGs possibly with self-loops), thereby limiting the number of
synchronisations between different threads. To this end, we may simply unfold
the \defn{underlying control-state graph} of the \sGTRS/ 
(see Section~\ref{def:underlying-control})
in the standard way, while preserving self-loops.
This type of underapproximation is similar to \emph{loop acceleration} in
the symbolic acceleration framework of \cite{BFLS05}.
Second, we bound the number of reversals between non-decrementing and
non-incrementing modes of the counters \cite{Iba78}.
Under these two restrictions, reachability is shown to be
NP-complete; in fact, it is poly-time reducible to satisfaction over existential
Presburger formulas, which allows one to tap into highly optimised SMT solvers.
Our result strictly generalises the decidability (in fact, NP-completeness) of
reachability for (i) weakly-synchronised ground-tree rewrite systems 
\cite{L12,TL10},
and
(ii) synchronisation/reversal-bounded concurrent pushdown systems with
counters \cite{HL12}.

Finally, we show one negative result that delineates the boundary of
decidability. If we relax the weak
control underapproximation by the notion of senescence (with age restrictions associated
with nodes in the trees) \cite{H14}, then the resulting model becomes
undecidable.


\paragraph{Related Work.}

Recursively-parallel program analysis was analysed in detail by Bouajjani and Emmi~\cite{BE13}.
However, in contrast to our systems, their model does not allow processes to communicate during execution.
Instead, processes hold handles to other processes which allow them to wait on the completion of others, and obtain the return value.
They show that when handles can be passed to child processes (during creation) then the state reachability problem is undecidable.
When handles may only be returned from a child to its parent, state reachability is decidable, with the complexity depending on which of a number of restrictions are imposed.

The work of Bouajjani and Emmi is closely related to branching vector addition systems~\cite{VGL05} which can model a stack of counter values which can be incremented and decremented (if they remain non-negative), but not tested.
While it is currently unknown whether reachability of a configuration is decidable, control-state reachability and boundedness are both 2ExpTime-complete~\cite{DJLL13}.

Another variant of vector addition systems with recursion are pushdown vector addition systems, where a single (sequential) stack and several global counters are permitted.
As before, these counters can be incremented and decremented, but not compared with a value.
Reachability of a configuration, and control-state reachability in these models remain open problems, but termination (all paths are finite) and boundedness are known to be decidable~\cite{LPS14}.
For reachability of a configuration, an under-approximation algorithm is proposed by Atig and Ganty where the stack behaviour is approximated by a \emph{finite index} context-free language~\cite{AG11}.

Lang and L\"oding study boundedness problems over sequential pushdown systems~\cite{LL13}.
In this model, the pushdown system is equipped with a counter that can be incremented, reset, or recorded.
Their model differs from ours first in the restriction to sequential systems, and second because the counter cannot effect execution or be decremented:
it is a recording of resource usage.
These kind of cost functions have also been considered over static trees~\cite{CL10,BCKPB14}, however, to our knowledge, they have not been studied over tree rewrite systems.


%% file: preliminaries.tex
\section{Preliminaries}
\label{sec:preliminaries}

We write $\N$ to denote the set of natural numbers and $\Z$ the set of integers.

\subsubsection{Trees}

A \emph{ranked alphabet} is a finite set of characters $\alphabet$ together with
a rank function $\rank : \alphabet \mapsto \N$.  A \emph{tree domain} $\tdom
\subset \N^\ast$ is a non-empty finite subset of $\N^\ast$ that is both
\emph{prefix-closed} and \emph{younger-sibling-closed}.  That is, if $\tnode
i \in \tdom$, then we also have $\tnode \in \tdom$ and, for all $1 \leq
j \leq i$, $\tnode j \in \tdom$ (respectively).  A \emph{tree} over
a ranked alphabet $\alphabet$ is a pair $\tree = \tup{\tdom, \tlab}$ where
$\tdom$ is a tree domain and $\tlab : \tdom \mapsto \alphabet$ such that for all
$\tnode \in \tdom$, if $\ap{\tlab}{\tnode} = \cha$ and $\ap{\rank}{\cha} = n$
then $\tnode$ has exactly $n$ children (i.e. $\tnode n \in \tdom$ and
$\tnode (n + 1) \notin \tdom$).  Let $\trees{\alphabet}$ denote the set of
trees over $\alphabet$.

\subsubsection{Context Trees}

A \emph{context tree} over the alphabet $\alphabet$ with a set of context
variables $\cvar_1, \ldots, \cvar_n$ is a tree $\context = \tup{\tdom,
\tlab}$ over $\alphabet \uplus \set{\cvar_1, \ldots, \cvar_n}$ such that for
each $1 \leq i \leq n$ we have $\ap{\rank}{\cvar_i} = 0$ and there
exists a unique \emph{context node} $\tnode_i$ such that
$\ap{\tlab}{\tnode_i} = \cvar_i$.  We will denote such a tree
$\csub{\context}{\cvar_1, \ldots, \cvar_n}$.
Given trees $\tree_i = \tup{\tdom_i, \tlab_i}$ for each $1 \leq
i \leq n$, we denote by $\csub{\context}{\tree_1, \ldots, \tree_n}$
the tree $\tree'$ obtained by filling each variable $\cvar_i$ with
$\tree_i$.  That is, $\tree' = \tup{\tdom', \tlab'}$ where
\[
    \tdom' = \tdom \cup \tnode_1 \cdot \tdom_1 \cup \cdots \cup \tnode_n \cdot
    \tdom_n
    \quad
    \text{and}
    \quad
    \ap{\tlab'}{\tnode} =
    \begin{cases}
        \ap{\tlab}{\tnode} & \tnode \in \tdom \land \forall i . \tnode \neq
        \tnode_i
        \\
        \ap{\tlab_i}{\tnode'} & \tnode = \tnode_i \tnode' \ .
    \end{cases}
\]

\subsubsection{Tree Automata}

A \emph{bottom-up non-deterministic tree automaton} (NTA) over a ranked alphabet
$\alphabet$ is a tuple $\ta = \tup{\tastates, \tarules, \tafinals}$ where
$\tastates$ is a finite set of states, $\tafinals \subseteq \tastates$ is a set
of final (accepting) states, and $\tarules$ is a finite set of rules of the form
$\tarule{\tast_1, \ldots, \tast_n}{\cha}{\tast}$ where $\tast_1, \ldots,
\tast_n, \tast \in \tastates$, $\cha \in \alphabet$ and $\ap{\rank}{\cha} =
n$.  A \emph{run} of $\ta$ on a tree $\tree = \tup{\tdom, \tlab}$ is a
mapping $\tarun : \tdom \mapsto \tastates$ such that for all $\tnode \in \tdom$
labelled $\ap{\tlab}{\tnode} = \cha$ with $\ap{\rank}{\cha} = n$ we have
$\tarule{\ap{\tarun}{\tnode 1}, \ldots, \ap{\tarun}{\tnode n}}
        {\cha}
        {\ap{\tarun}{\tnode}}$.
It is accepting if $\ap{\tarun}{\eword} \in \tafinals$.  The \emph{language}
defined by a tree automaton $\ta$ over alphabet $\alphabet$ is a set
$\ap{\lang}{\ta} \subseteq \trees{\alphabet}$ of trees over which there exists 
an accepting run of $\ta$.

\subsubsection{Parikh images}

Given an alphabet
$\alphabet = \{\osym_1,\ldots,\osym_n\}$ and a word $w \in
\alphabet^*$, we write $\Parikh{w}$ to denote a mapping $\lambda: \alphabet
\to \N$, where $\lambda(a)$ is defined to be the number of occurrences of
$a$ in $w$. Given a language $L \subseteq \alphabet^*$, we write
$\Parikh{L}$ to denote the set $\setcomp{\Parikh{w}}{w \in L}$. We say that
$\Parikh{L}$ is the \defn{Parikh image} of $L$.

\subsubsection{Presburger Arithmetic}

Presburger formulas are first-order formulas over integers with addition.
Here, we use extended existential Presburger formulas $\varphi(\vecX,\vecY) :=
\exists \vecX \varphi$, where (i) $\vecX$ and $\vecY$ are sets of variables,
and (ii) $\varphi$ is a boolean combination of expressions $\sum_{i=1}^m a_iz_i 
\sim b$ for variables $z_1,\ldots,z_m \in \vecX \cup \vecY$,
constants $a_1,\ldots,a_m,b\in \Z$, and $\sim\ \in \{\leq,\geq,<,>,=\}$ with 
constants represented in binary. A \defn{solution} to $\varphi$ is a 
valuation $\vecB: \vecY \mapsto \Z$ to $\vecY$ such that 
$\varphi(\vecX,\vecB)$ is true. The formula $\varphi$ is \defn{satisfiable} if 
it
has a solution. Satisfiability of existential Presburger formulas is NP-complete even with
these extensions (cf.  \cite{Sca84}).

%% file: model.tex
\section{Formal Models}
\label{sec:models}

In this section, we will define our formal models, which are based on
ground-tree rewrite systems.
Ground-tree rewrite systems (GTRSs) \cite{DT90} permit subtree rewriting where
rules are given as a pair of ground-trees. In the sequel, we use the extension
proposed by L\"{o}ding \cite{L06} where NTA (instead of
ground
trees) appear in the rewrite rules. Hence, a single rule may correspond to an
infinite number of \defn{concrete rules} (i.e. containing concrete trees).

\subsubsection{Ground Tree Rewrite Systems with State and Reversal Bounded
Counters.}

To capture synchronisations between different subthreads, we follow
\cite{L12,KRS04,TL10} and extend GTRS with state (a.k.a. global control).
The resulting model is denoted by \sGTRS/ (state-extended GTRS).
To capture integer variables, we further extend the model with unbounded
integer counters, which can
be incremented, decremented, and compared against an integer constant. Since
Minsky's machines can easily be encoded in such a model, we apply
a standard underapproximation technique: \emph{reversal-bounded analysis of
the counters} \cite{HL11,Iba78}. This means that one only analyses executions
of the machines whose number of reversals between nondecrementing and
nonincrementing modes of the counters is bounded by a given constant $r
\in \N$ (represented in unary). The resulting model will be denoted by \rbGTRS/.
We will now define this model in more detail.

\OMIT{
An \rbGTRS/ maintains a tree over a given alphabet $\alphabet$ and a control state from a finite set.
Each transition may update the control state and rewrites a part of the tree.
Rewriting a tree involves matching a sub-tree of the current tree and replacing it with a new tree.
Note, that since we are considering ranked trees, a sub-tree cannot be erased by a rewrite rule, since this would make the tree inconsistent w.r.t the ranks of the tree labels.
\al{Remove this or put the last sentence after precise definition}
}

An \defn{atomic counter constraint} on counter variables
$\ctrs = \{\ctr_1,\ldots,\ctr_k\}$
is an expression of the form
$\ctr_i \compop \val$,
where
$\val \in \Z$
and
$\compop \in \set{<,\leq,=,\geq,>}$.
A
\defn{counter constraint} $\con$ on $\ctrs$ is a boolean combination of atomic counter constraints on $\ctrs$.
Given a valuation
$\vals : \ctrs \mapsto \Z$
to the counter variables, we can determine whether
$\con[\nu]$
is true or false by replacing a variable $\ctr$ by $\nu(\ctr)$ and evaluating the resulting arithmetic expressions in the obvious way.
Let $\ccons{\ctrs}$
denote the set of all counter constraints on $\ctrs$.
Intuitively, these formulas will act as guards to determine whether certain transitions can be fired.
Given two counter valuations $\vals$ and $\dvals$ we define
$\vals + \dvals$
as the pointwise addition of the valuations.
That is,
$(\vals + \dvals)(\ctr) = \vals(\ctr) + \dvals(\ctr)$.

Given a sequence of counter values, a reversal is when a counter switches from being incremented to be decremented or vice-versa.
For example, if the values of a counter $\ctr$ along a run are
$1,1,1,2,3,4,4,\overline{4,3},2,\overline{2,3}$,
then the number of reversals of $\ctr$ is 2
(reversals occur in between the overlined positions).
A sequence of valuations is reversal-bounded whenever the number of reversals is the sequence is bounded.

\begin{definition}[$r$-Reversal-Bounded]
    For a counter $\ctr$ from a set of counters $\ctrs$, a sequence
    $\vals_1, \ldots, \vals_n$
    of counter valuations over $\ctrs$ is \emph{$r$-reversal-bounded} for $\ctr$ whenever we can partition
    $\vals_1, \ldots, \vals_n$
    into $(r+1)$ sequences
    $\phase_1, \ldots, \phase_{r+1}$
    (with
    $\vals_0, \ldots, \vals_n
     =
     \phase_1, \ldots, \phase_{r+1}$)
    such that for all
    $1 \leq i \leq r$
    there is some
    $\compop \in \set{\leq, \geq}$
    such that for all
    $\vals_j, \vals_{j+1}$
    appearing together in $\phase_i$, we have
    $\ap{\vals_j}{\ctr} \compop_\ctr \ap{\vals_{j+1}}{\ctr}$.
\end{definition}

We define \sGTRS/ with reversal-bounded counters.

\begin{definition}[\sGTRS/s with $r$-Reversal-Bounded Counters (\rbGTRS/)]
    A \emph{state-extended ground tree rewrite system with $r$-reversal-bounded
    counters} (\rbGTRS/) is a tuple
    $\gtrs = \tup{\controls, \alphabet, \oalphabet, \rules, \ctrs, r}$
    where
        $\controls$ is a finite set of control-states,
        $\alphabet$ is a finite ranked alphabet,
        $\oalphabet$ is a finite alphabet of output symbols (i.e. transition
        labels),
        $\rules$ is a finite set of rules of the form
            $\ctrrule{\control_1}{\ta_1}{\con}
                     {\osym}
                     {\control_2}{\ta_2}{\dvals}$
            where
                $\control_1, \control_2 \in \controls$,
                $\osym \in \oalphabet$,
                $\con \in \ccons{\ctrs}$,
                $\dvals \in \ctrs \mapsto \Z$, and
                $\ta_1, \ta_2$
                    are NTAs over $\alphabet$.
\end{definition}
In the sequel, we will omit mention of the number $r$ in the tuple
$\gtrs$ if it is clear from the context.

A \defn{configuration} of an \sGTRS/ with counters is a tuple
$\conf = \ctrconfig{\control}{\tree}{\vals}$
where $\control$ is a control-state, $\tree$ a tree, and $\vals$ a valuation of the counters.
We have a \emph{transition}
$\ctrconfig{\control_1}{\tree_1}{\vals_1}
 \tran{\osym}
 \ctrconfig{\control_2}{\tree_2}{\vals_2}$
whenever there is a rule
$\ctrrule{\control_1}{\ta_1}{\con}{\osym}{\control_2}{\ta_2}{\dvals} \in \rules$
such that: (i) \defn{(dynamics of counters)} $\con[\vals_1]$ is true and
$\vals_2 = \vals_1 + \dvals$, and (ii) \defn{(dynamics of trees)} $\tree_1 =
\csub{\context}{\tree'_1}$ for some context $\context$ and tree
$\tree'_1 \in \ap{\lang}{\ta_1}$ and
$\tree_2 = \csub{\context}{\tree'_2}$
for some tree $\tree'_2 \in \ap{\lang}{\ta_2}$.
A \emph{run} $\gtrsrun$ over
$\osym_1\ldots\osym_{n-1}$
is a sequence
\[
    \ctrconfig{\control_1}{\tree_1}{\vals_1} \tran{\osym_1} \cdots
    \tran{\osym_{n-1}} \ctrconfig{\control_n}{\tree_n}{\vals_n}
\]
such that for all
$1 \leq i < n$
we have
$\ctrconfig{\control_i}{\tree_i}{\vals_i}
 \tran{\osym_i}
 \ctrconfig{\control_{i+1}}{\tree_{i+1}}{\vals_{i+1}}$
is a transition of $\gtrs$ and
for each
$\ctr \in \ctrs$
the sequence
$\vals_1, \ldots, \vals_n$
is $r$-reversal-bounded for $\ctr$.
We say that $\osym_1\ldots\osym_{n-1}$ is the output string of $\gtrsrun$.
We write
$\ctrconfig{\control}{\tree}{\vals}
 \run{\osym_1\ldots\osym_n}
 \ctrconfig{\control'}{\tree'}{\vals'}$
 (or simply
$\ctrconfig{\control}{\tree}{\vals}
 \to^*
 \ctrconfig{\control'}{\tree'}{\vals'}$)
whenever there is a run from
$\ctrconfig{\control}{\tree}{\vals}$
to
$\ctrconfig{\control'}{\tree'}{\vals'}$
over
$\osym_1\ldots\osym_n$.
Let $\empsym$ denote the empty output symbol.

Whenever we wish to discuss \sGTRS/s without counters, we simply omit the counter components.
That is, we have configurations of the form
$\config{\control}{\tree}$
and transitions of the form
$\rrule{\control_1}{\ta_1}{\osym}{\control_2}{\ta_2}$.
The standard notion of GTRS (i.e. not state-extended) \cite{L06} is simply
\sGTRS/ without counters with only one state.

We next define the problems of \defn{(global) reachability}.
To this end, we use a tree automaton $\ta$ (resp. an existential
Presburger formula $\varphi$) to represent the tree (resp. counter) component
of a configuration. More precisely, a \defn{symbolic config-set}
of an \rbGTRS/
$\gtrs = \tup{\controls, \alphabet, \oalphabet, \rules, \ctrs, r}$
is a tuple $\ctrconfig{\control}{\ta}{\varphi}$, where $\control \in \controls$,
$\ta$ is an NTA over $\alphabet$, and $\varphi(\bar x)$ is an existential
Presburger formula with free variables $\bar x = \{x_\ctr\}_{\ctr \in \ctrs}$
(i.e. one free variable
for each counter). Each symbolic config-set $\ctrconfig{\ctr}{\ta}{\varphi}$ represents
a set of configurations of $\gtrs$ defined as follows:
$\sem{\ctrconfig{\control}{\ta}{\varphi}} :=
\{ \ctrconfig{\control}{\tree}{\vals} : \tree \in \ap{\lang}{\ta},
\text{ $\varphi(\vals)$ is true} \}$.
\problemx{Global Reachability}
         {an \rbGTRS/ $\gtrs$ and two symbolic config-sets
            $\ctrconfig{\control_1}{\ta_1}{\varphi_1}$
            $\ctrconfig{\control_2}{\ta_2}{\varphi_2}$}
         {Decide whether
            $\ctrconfig{\control_1}{\tree_1}{\vals_1} \to^*
            \ctrconfig{\control_2}{\tree_2}{\vals_2}$, for some
            $\ctrconfig{\control_1}{\tree_1}{\vals_1} \in
             \sem{\ctrconfig{\control_1}{\ta_1}{\varphi_1}}$ and
            $\ctrconfig{\control_2}{\tree_2}{\vals_2} \in
             \sem{\ctrconfig{\control_2}{\ta_2}{\varphi_2}}$
         }
\noindent
The problem of \defn{control-state reachability} can be defined by restricting
(i) the tree automata $\ta_1$ and $\ta_2$ to accept, respectively, a singleton
tree and the set of all trees, and (ii) the solutions to the formulas 
$\varphi_1$ and $\varphi_2$ are, respectively, $\{\valszero\}$ (where
$\valszero$ is the valuation assigning $0$ to all counters) and the set of all
counter valuations.
\smallskip

\begin{remark}
    When we measure the complexity of reachability for \rbGTRS/, the number $r$
    of reversals is represented in unary, while the numbers in counter
    constraints and valuations are represented in binary. This is consistent
    with the standard representation of numbers in previous work on
    reversal-bounded counter machines (e.g. see \cite{HL11,HL12}).
\end{remark}

\subsubsection{Weakly Synchronised Ground Tree Rewrite Systems}

The control-state and global reachability problems for \sGTRS/ are known to be
undecidable~\cite{BKRS09,GL11}.
The problems become NP-complete for \emph{weakly-synchronised}
\sGTRS/~\cite{L12,TL10}, where the underlying control-state graph (where there is an edge between $\control_1$ and $\control_2$ whenever there is a transition
$\rrule{\control_1}{\ta_1}{\osym}{\control_2}{\ta_2}$)
may only have cycles of length $1$ (i.e. self-loops), i.e., a DAG (directed
acyclic graph) possibly with self-loops. Underapproximation by a weak control
is akin to loop acceleration in the symbolic acceleration framework of
\cite{BFLS05}.
We extend the definition to \rbGTRS/s.
The original definition can be easily obtained by omitting the counter components.

\label{def:underlying-control}
We define the \emph{underlying control graph} of an \rbGTRS/
$\gtrs = \tup{\controls, \alphabet, \oalphabet, \rules, \ctrs}$
as a tuple
$\tup{\controls, \nfatrans}$
where
    $\nfatrans = \setcomp{\tup{\control_1, \control_2}}
                        {\ctrrule{\control_1}{\ta_1}{\con}
                                 {\osym}
                                 {\control_1}{\ta_2}{\dvals} \in \rules} \ .$

\begin{definition}[Weakly-Synchronised \rbGTRS/]
    An \rbGTRS/ is \emph{weakly synchronised} if its underlying control graph
    $\tup{\controls, \nfatrans}$
    is a DAG possibly with self-loops.
    \OMIT{
    such that all paths
    \[
        \tup{\control_1, \control_2}\tup{\control_2,
        \control_3}\ldots\tup{\control_{n-2},
        \control_{n-1}}\tup{\control_{n-1}, \control_n} \in
        \nfatrans^\ast
    \]
    with
    $\control_1 = \control_n$
    satisfy
    $\control_i = \control_1$
    for all
    $1 \leq i \leq n$.
}
\end{definition}

%% file: weak.tex
\section{Decidability}
\label{sec:weak}

In this section we will prove the main result of the paper:
\begin{theorem}
Global reachability for weakly synchronised \rbGTRS/ is NP-com-\\plete. In fact,
    it is poly-time
    reducible to satisfiability over existential Presburger formulas.
    \label{th:main}
\end{theorem}
To prove this theorem, we fix notation for the input to the
problem: an \rbGTRS/
$\gtrs = \tup{\controls, \alphabet, \oalphabet, \rules, \ctrs, r}$
and two symbolic config-sets $\ctrconfig{\control_1}{\ta_1}{\varphi_1}$,
$\ctrconfig{\control_2}{\ta_2}{\varphi_2}$ of $\gtrs$. Let $\ctrs =
\{\ctr_i\}_{i=1}^k$. The gist of the
proof is as follows. From $\gtrs$, we construct a new \sGTRS/
$\gtrs'$ (without counters) by encoding the dynamics of the counters in the
output symbols of $\gtrs'$. Of course, $\gtrs'$ has no way of comparing the
values of counters with constants. [In this sense, $\gtrs'$ only
overapproximates the behavior of $\gtrs$.] To deal with this problem, we use the
result of \cite{L12} to compute an existential Presburger formula $\psi$
capturing the Parikh images of the set of all output strings of $\gtrs'$
from $\ctrconfig{\control_1}{\ta_1}{\varphi_1}$ to
$\ctrconfig{\control_2}{\ta_2}{\varphi_2}$. The final
formula is $\psi \wedge \psi'$, where $\psi$ is a constraint asserting that
the desired counter comparisons are performed throughout runs of $\gtrs'$.
We sketch the details of the construction below.

\paragraph{Modes of the counters.} The first notion that is crucial in our
proof is that of \emph{mode} of a counter \cite{HL11,Iba78}, which is an
abstraction of the values of a counter
in a run of an \rbGTRS/ containing three pieces of information:
(i) the \emph{region} of the counter value (i.e. how it compares to constants
occurring in counter constraints), (ii) the number of reversals that
has been performed by each counter (between 0 and $r$), and (iii) whether a counter is
currently non-decrementing ($\uparrow$) or non-incrementing ($\downarrow$).
A \defn{mode vector} is simply a $k$-tuple of modes, one mode for each
of the $k$ counters. We now formalise these notions.

Let $d_1 < \ldots < d_m$ be the integer constants appearing in the counter
constraints in $\gtrs$. This sequence of constants gives rise to the set
$\REG$ of \defn{regions} defined as $\REG := \{ \regA_0,\ldots,\regA_m,
\regB_1,\ldots,\regB_m\}$, where $\regB_i := \{d_i\}$ (where $1 \leq i \leq m$),
$\regA_i := \{ n \in \Z: d_i < n < d_{i+1} \}$ (where $1 \leq i < m$),
$\regA_0 := \{ n \in \Z: n < d_1 \}$, and
$\regA_m := \{ n \in \Z: n > d_m \}$. A \defn{mode} is simply a tuple in
$\REG \times [0,r] \times \{\uparrow,\downarrow\}$. A \defn{mode vector} is
simply a tuple in $\Modes := \REG^k \times [0,r]^k \times
\{\uparrow,\downarrow\}^k$.

\paragraph{Building the \sGTRS/ $\gtrs'$.}
We might be tempted to build $\gtrs'$ by first removing the counters from
$\gtrs$ and then embedding $\Modes$ into
the control states $\gtrs'$. This, however, causes two problems. First, the
number of control states becomes exponential in $k$. Second, the resulting
system is no longer weakly synchronised even though $\gtrs$ originally was
weakly synchronised. To circumvent this problem, we apply adapt a technique
from \cite{HL11}. Every run $\gtrsrun$ of $\gtrs$ from
$\ctrconfig{\control_1}{\ta_1}{\varphi_1}$
to $\ctrconfig{\control_2}{\ta_2}{\varphi_2}$ can be associated with a
sequence $\sigma$ of mode vectors recording the information (i)--(iii) for each
counter. The crucial observation is that there are at most $N_{\max} :=
2mk(r+1)$ different
modes in $\sigma$. This is because a counter can only go through at most
$2m$ regions without incurring a reversal. For this reason, we may use the
control states of $\gtrs'$ to store the number of mode vectors that
$\gtrs$ has gone through, while the actual mode vector guessed by
$\gtrs'$ will be made ``visible'' in the output strings of $\gtrs'$. That way,
we can use an additional existential Presburger formula $\psi'$ (see
below) to enforce that the run of $\gtrs'$ faithfully simulates runs of
$\gtrs$. In addition, the shape of the control states (DAG with self-loops) of
$\gtrs'$ is preserved. [The product graph of two DAGs with self-loops is
also a DAG with self-loops.] We detail the construction below.

Define the weakly-synchronised \sGTRS/ $\gtrs' = \tup{\controls', \alphabet,
\oalphabet', \rules'}$ as follows. Let $\controls' := \controls \times
[0,N_{\max}]$. The output alphabet $\oalphabet'$ is defined as
$\oalphabet \times \rules \times [0,N_{\max}] \times \{0,1\}$, where the
boolean flag is used to denote whether the transition taken changes the mode.
We define $\rules'$ as follows. For
each rule
            $\rruler = \ctrrule{\control}{\ta}{\con}
                     {\osym}
                     {\control'}{\ta'}{\dvals}$
in $\rules$, we add the rule
            $\rrule{(\control,i)}{\ta}{(\osym,\rruler,i,0)}{(\control',i)}{\ta'}$
for each $i \in [0,N_{\max}]$, and
         $\rrule{(\control,i)}{\ta}{(\osym,\rruler,i,1)}{(\control',i+1)}{\ta'}$
for each $i \in [0,N_{\max})$.
\OMIT{
We also add a bunch of rules that can \emph{instantaneously} increment the
mode counter in the states:
\mh{It seems strange that we can have these rules but also need to output a boolean when we explicitly change modes}
$\rrule{(\control,i)}{\ta}{\epsilon}{(\control,i+1)}{\ta}$, for each
$\control \in \controls$, each $i \in [0,N_{\max})$, and each NTA
$\ta$ (up to language equivalence) recognising a singleton language containing
a single-node tree. [In other words, the tree component is unconstrained, but
has to remain unchanged under transitions.]
}
Since $\gtrs$ is
weakly-synchronised and the mode counter never decreases, it follows that
$\gtrs'$ is weakly-synchronised too. Note also that this construction can
be performed in polynomial-time.

\paragraph{Constructing the formula $\psi \wedge \psi'$.}
As we mentioned, $\psi$ is an existential Presburger formula encoding the
Parikh image $\Parikh{L}$ of the set $L$ of all output strings of $\gtrs'$ from
$\config{(\control_1,0)}{\ta_1}$ to
$\config{S}{\ta_2}$, where
$S = \{\control_2\} \times [0,N_{\max}]$.
More precisely, the set $\vecZ$ of
free variables of $\psi$ include $z_a$ for each $a \in \oalphabet'$.
Furthermore, for each valuation $\dvals \in \vecZ \mapsto \Z$, it is the case
that $\psi(\dvals)$ is true iff $\dvals \in \Parikh{L}$. Such a formula is
known to be polynomial-time computable since $\gtrs'$ is a weakly-synchronised
\sGTRS/ \cite{L12}.

Recall that $\psi'$ should assert that the desired counter comparisons are
performed throughout runs of $\gtrs'$. To this end, the formula $\psi'$ will
have extra variables for guessing the existence of a sequence of $N_{\max}$
distinct mode vectors through
runs of $\gtrs'$. More precisely, the formula $\psi'$ is the conjunction
\begin{center}
\begin{tabular}{l}
    $\varphi_1(\vecX) \wedge \varphi_2(\vecY) \wedge
    \Dom(\vecM_0,\ldots,\vecM_{N_{\max}}) \wedge \Init(\vecM_0) \wedge$\\
    $\GoodSeq(\vecM_0,\ldots,\vecM_{N_{\max}}) \wedge
    \Respect(\vecZ,\vecM_0,\ldots,\vecM_{N_{\max}}) \wedge
        \EndVal(\vecX,\vecY,\vecZ)$.
\end{tabular}
\end{center}
\noindent
The set $\vecX$ consists of variables $x_i$ ($1 \leq i \leq k$)
which contain the initial value of the $i$th counter.
Similarly, the set $\vecY$ consists of variables $y_i$ ($1 \leq i
\leq k$) which contain the final value of the $i$th counter.
Each $\vecM_i$ denotes a set of variables for the $i$th mode vector defined
as follows:
\begin{itemize}
    \item $\regVar{i}{j}$ (for each $j \in [1,k]$) --- to encode which of
        the $2m+1$ possible regions the $j$th counter is in.
    \item $\revVar{i}{j}$ (for each $j \in [1,k]$) --- to encode how many
        reversals have been used up by the $j$th counter.
    \item $\arrVar{i}{j}$ (for each $j \in [1,k]$) --- to encode whether
        the $j$th counter is non-incrementing or non-decrementing.
\end{itemize}
We detail each subformula below.

The subformula $\Dom$ asserts that each variable in $\vecM_i$ (for each $i$)
has the right domain (i.e. range of integer values). More precisely, for each
$j \in [1,k]$, we add the conjuncts: (i) $0 \leq \regVar{i}{j} \leq 2m$,
(ii) $0 \leq \revVar{i}{j} \leq r$, and (iii) $0 \leq \arrVar{i}{j} \leq 1$.
For the first constraint, we use an even number of the form $2i$ to represent
the region $A_i$, and an odd number $2i-1$ to represent the region $B_i$.
The last constraint simply encodes non-decrementing ($\uparrow$) as 1, and
non-incrementing ($\downarrow$) as 0.

The subformula $\Init$ asserts that $\vecM_0$ is an initial mode vector. More
precisely, for each $j \in [1,k]$, we add the conjuncts $\revVar{0}{j}
= 0$.

The subformula $\GoodSeq$ asserts that $\vecM_0,\ldots,\vecM_{N_{\max}}$
forms a valid sequence of mode vectors. More precisely, for each $i \in
[0,N_{\max})$ and each $j \in [1,k]$, we add the conjuncts: (i)
$\arrVar{i}{j} \neq \arrVar{i+1}{j}
\Implies \revVar{i+1}{j} = \revVar{i}{j}+1$, (ii) $\arrVar{i}{j} =
\arrVar{i+1}{j} \Implies \revVar{i+1}{j} = \revVar{i}{j}+1$, (iii)
$\regVar{i}{j} < \regVar{i+1}{j} \Implies \arrVar{i+1}{j} = 0$, and
(iv) $\regVar{i}{j} > \regVar{i+1}{j} \Implies \arrVar{i+1}{j} = 1$.
For example, the first constraint asserts that a change in the direction
(non-incrementing or non-decrementing) of the counter incurs one reversal.
The other constraints are similar.

The subformula $\Respect$ asserts that the Parikh image $\vecZ$ of the run of
$\gtrs'$ respects the sequence $\vecM_0,\ldots,\vecM_{N_{\max}}$ of mode
vectors. In effect, this subformula ensures that $\gtrs'$ faithfully simulates
$\gtrs$. Firstly, we need to assert that the $j$th counter values at the
\emph{start} and at the \emph{end} of the $i$th mode of $\gtrs'$ (which are
encoded in $\vecZ$) are in the right regions $\regVar{i}{j}$. To
state this more precisely, for each rule
            $\rruler = \ctrrule{\control}{\ta}{\con}
                     {\osym}
                     {\control'}{\ta'}{\dvals}$
in $\rules$, we let $\mu_j(\rruler)$ denote the value $\dvals(c_j)$.
For each $i \in [0,N_{\max}]$ and $j \in [1,k]$, we denote
by the notation $\StartCounter{i}{j}$ the term
    $x_j + \sum_{s=0}^{i-1}\sum_{(\osym,\rruler,s,l)} \mu_j(\rruler) \times
                z_{(\osym,\rruler,s,l)}$,
where $\osym$, $\rruler$, and $l$, range over, respectively, $\oalphabet$,
$\rules$, and $\{0,1\}$. Similarly, we denote by  $\EndCounter{i}{j}$
the term $\StartCounter{i}{j} + \sum_{(\osym,\rruler,i,0)} \mu_j(\rruler) \times
z_{(\osym,\rruler,i,0)}$. We add the conjuncts:
(i) $\regVar{i}{j} = 2h \Implies \EndCounter{i}{j} \in A_h$, for each
$h \in [0,m]$, and
(ii) $\regVar{i}{j} = 2h+1 \Implies \EndCounter{i}{j} \in B_h$, for each
$h \in [0,m)$. 
[Note that formulas of the form $g \in A$, for a Presburger
term $g$ and a set $S \in \{A_0,\ldots,A_m,B_1,\ldots,B_m\}$, can be easily
replaced by quantifier-free Presburger formulas, e.g., $g \in A_0$ stands for
$g < d_1$.] 
To ensure that the initial condition is correct, for each $j \in [1,k]$,
we add the following conjuncts: (1) $\StartCounter{0}{j} \in A_h
\Implies \regVar{0}{j} = 2h$, and (2) $\StartCounter{0}{j} \in B_h
\Implies \regVar{0}{j} = 2h+1$.
Secondly, we need to state that the transitions executed in
each mode are valid (i.e. satisfy the counter constraints). More precisely,
for each $\osym \in \oalphabet$, $\rruler \in \rules$, $i \in [0,N_{\max}]$,
and $l \in \{0,1\}$, if $\con$ is the counter constraint in $\rruler$, we
add the conjunct $z_{(\osym,\rruler,i,l)} > 0 \Implies
\con(\StartCounter{i}{1},\ldots,\StartCounter{i}{k})$. Finally, we assert
that, when the $j$th counter is non-incrementing (resp. non-decrementing), only
non-negative (resp. non-positive) counter increments are permitted. More
precisely, for each $i \in [0,N_{\max}]$, $j \in [1,k]$, $l \in \{0,1\}$, and
$\rruler \in \rules$, if $\mu_j(\rruler) > 0$, then add the conjunct
$\arrVar{i}{j} = 0 \Implies z_{(\osym,\rruler,i,l)} = 0$; if $\mu_j(\rruler) < 0$, then
add the conjunct $\arrVar{i}{j} = 1 \Implies z_{(\osym,\rruler,i,l)} = 0$.

Finally, the subformula $\EndVal$ simply asserts that, starting from the initial
counter value $\vecX$ and following the transitions $\vecZ$, the end counter
values are $\vecY$. To this end, we can simply add the conjunct
$y_j = \EndCounter{N_{\max}}{j}$ for each $j \in [1,k]$.

This concludes the formula construction. It is immediate that $\gtrs'$
faithfully simulates $\gtrs$ iff $\psi \wedge \psi'$ is true.
    %
In addition, the formula construction
runs in polynomial-time. Since satisfiability over existential
Presburger formulas is NP-complete \cite{Sca84}, the NP upper bound for Theorem
\ref{th:main} follows. NP-hardness already holds for the restricted model
where the tree component is a stack \cite{HL11}.

%% file: senescent.tex
\section{Senescent Ground-Tree Rewrite Systems}
\label{sec:senescent}

In this section we relax the weakly-extended restriction to the notion of \emph{senescence}~\cite{H14}.
Due to limited space, we relegate formal proofs and definitions to the
\shortlong{full version}{appendix}.
We show the following result.

\begin{namedtheorem}{thm:senescent-undec}{Control State Reachability of Senescent \rbGTRS/}
    The control-state reachability problem for senescent \rbGTRS/ is undecidable.
\end{namedtheorem}

Senescence allows the underlying control-state graph to have arbitrary cycles (instead of only self-loops).
For \sGTRS/, control-state reachability is decidable under an ``age restriction'' that is imposed on the nodes that can be rewritten.
That is, when the control-state changes, the nodes in the tree age by one timestep.
Once a node reaches an \textit{a priori} fixed age $\agebound$, it becomes
fixed (i.e. cannot be rewritten by further transitions in the run).

We show control-state reachability for senescent \rbGTRS/s is undecidable.
In the following, we refer to nodes whose age is within the age bound as \emph{live}.
Note, each time a node is rewritten, its age is reset to zero and we can keep leaves of the tree live by allowing them to rewrite to themselves.

We follow the proof that reachability for reset Petri nets is undecidable~\cite{AK77}.
We simulate a two-counter machine.
Testing whether such a machine can reach a given control-state while having counters with value zero is undecidable.

In the tree, we track the value of a counter by the number of live leaves labelled with the counter name.
To increment a counter we add a new leaf.
To decrement a counter, we rewrite a leaf to a null label.
We also track, using reversal-bounded counters, the number of increments made to each counter, and in separate counters, the number of decrements.
These are needed to ensure accuracy of the zero tests, which are simulated as follows.

To simulate a zero test, we perform the following checks.
First, we ``reset'' the counter to zero by forcing enough control-state changes to fix the nodes corresponding to the counter.
After this operation, the counter value is zero.
However, if the counter was not zero before the test, there will be a discrepancy with the reversal bounded counters: more increments will be recorded than decrements.
This cannot be corrected by the simulation.
Thus, at the end of the run, we check whether the number of increments is equal to the number of decrements.
If not, we know the run made a spurious transition.
If it is, we know the two-counter machine has a corresponding run.
This completes the gist of the simulation of a two-counter machine.

%% file: conclusion.tex
\section{Extensions and Future Work}
\label{sec:future}

We proposed \sGTRS/ with counters as a model of recursively parallel programs with unbounded recursion, thread creation, and integer variables.
To obtain decidability, we gave an underapproximation in the form of weak \sGTRS/ with reversal-bounded counters.
We showed that the reachability problem for this model is NP-complete; in fact,
polynomial-time reducible to satisfiability of linear integer arithmetic, for
which highly optimised SMT solvers are available (e.g. Z3 \cite{Z3}).
Additionally, we explored the possibility of relaxing the weakly-synchronised 
constraint to that of senescence, and showed that the resulting model has an undecidable control-state reachability problem.

One possible avenue of future work is to investigate what happens when \emph{local} integer values are permitted.
That is, reversal-bounded counters can be stored on the nodes of the tree.
We may also study techniques that allow nodes to contain multiple labels, permitting the modelling of multiple local variables without an immediate exponential blow up.

%% file: senescent-appendix.tex
\section{Proofs and Definitions for Senescent \sGTRS/}

We first give the definition of senescent \rbGTRS/s before giving the formal undecidability proof.

\subsection{Model Definition}

Given a run
\[
    \ctrconfig{\control_1}{\tree_1}{\vals_1}
    \tran{\osym_1}
    \cdots
    \tran{\osym_{n-1}}
    \ctrconfig{\control_n}{\tree_n}{\vals_n}
\]
of an \rbGTRS/, let
$\context_1, \ldots, \context_{n-1}$
be the sequence of tree contexts used in the transitions from which the run was constructed.
That is, for all
$1 \leq i < n$,
we have
$\tree_i = \csub{\context_i}{\treeout_i}$
and
$\tree_{i+1} = \csub{\context_i}{\treein_{i+1}}$
where
$\ctrrule{\control_i}{\ta_i}{\con_i}{\osym_i}{\control_{i+1}}{\ta'_i}{\dvals_i}$
was the rewrite rule used in the transition and
$\treeout_i \in \ap{\lang}{\ta_i}$,
$\treein_{i+1} \in \ap{\lang}{\ta'_i}$
were the trees that were used in the tree update.

For a given position
$\ctrconfig{\control_i}{\tree_i}{\vals_i}$
in the run and a given node $\tnode$ in the domain of $\tree_i$, the \emph{birthdate} of the node is the largest
$1 \leq j \leq i$
such that $\tnode$ is in the domain of
$\csub{\context_j}{\treein_j}$
and $\tnode$ is in the domain of
$\csub{\context_j}{\cvar}$
only if its label is $\cvar$.
The \emph{age} of a node is the cardinality of the set
$\setcomp{i'}{j \leq i' < i \land \control_{i'} \neq \control_{i'+1}}$.
That is, the age is the number of times the control-state changed between the $j$th and the $i$th configurations in the run.

A lifespan-restricted run with a lifespan of $\agebound$ is a run such that each
transition
$\ctrconfig{\control_i}{\csub{\context_i}{\treeout_i}}{\vals_i}
 \tran{\osym_i}
 \ctrconfig{\control_{i+1}}{\csub{\context_i}{\treein_{i+1}}}{\vals_{i+1}}$
has the property that all nodes $\tnode$ in $\treeout_i$ have an age of at most
$\agebound$.
That is, more precisely, that all nodes $\tnode$ in the domain of
$\csub{\context_i}{\treeout_i}$
but only in the domain of
$\csub{\context_i}{\cvar}$
if the label is $\cvar$ have an age of at most $\agebound$.

\begin{definition}[Senescent \rbGTRS/]
    A \emph{senescent \rbGTRS/} with \emph{lifespan} $\agebound$ is an \rbGTRS/
    $\gtrs = \tup{\controls, \alphabet, \rules, \ctrs}$
    where runs are \emph{lifespan-restricted} with a lifespan of $\agebound$.
\end{definition}

\OMIT{
\al{
    I was rather confused by the formal definition, so I thought I put some
    intuition. Let me know if the following is good to put.
}
\mh{Added a variant}
}
Note that
the senescence restriction is weaker than the weakly-synchronised
restriction in that the number of times the finite control could
change state is unbounded. In fact, a node could be affected by
an unbounded number of control state changes so long as it is always rewritten
without becoming fixed (i.e. reaches age $\agebound$).

\subsection{Undecidability}

We show that the control-state reachability problem
\OMIT{\mh{or whatever we call it}}
is undecidable via a reduction from the reachability problem for two-counter machines.

A two-counter machine is a tuple
$\cm = \tup{\cmcontrols, \cmrules}$
where
    $\controls$ is a finite set of control-states,
    $\cmrules$ is a finite set of rules of the form
        $\cmrule{\control_1}{\cmop}{\control_2}$
        where
            $\control_1, \control_2 \in \cmcontrols$, and
            $\cmop \in \set{\cminc{1}, \cminc{2},
                            \cmdec{1}, \cmdec{2},
                            \cmzero{1}, \cmzero{2}}$.
A configuration of $\cm$ is a tuple
$\cmconfig{\control}{\val_0}{\val_1} \in \cmcontrols \times \N \times \N$.
We have a transition
$\cmconfig{\control_1}{\val^1_0}{\val^1_1}
 \cmtran
 \cmconfig{\control_2}{\val^2_0}{\val^2_1}$
if we have a rule
$\cmrule{\control_1}{\cmop}{\control_2}$
and
\begin{itemize}
\item
    if
    $\cmop = \cminc{i}$,
    $\val^2_i = \val^1_i + 1$
    and
    $\val^2_{1-i} = \val^1_{1-i}$,
\item
    if
    $\cmop = \cmdec{i}$,
    $\val^2_i = \val^1_i - 1 \geq 0$
    and
    $\val^2_{1-i} = \val^1_{1-i}$,
\item
    if
    $\cmop = \cmzero{i}$,
    $\val^2_i = \val^1_i = 0$,
    and
    $\val^2_{1-i} = \val^1_{1-i}$.
\end{itemize}

Let $\valszero$ be the valuation assigning $0$ to all counters.
For given two-counter machine $\cm$ and control-states $\cmcontrol_0$ and $\cmcontrol_f$ we define a senescent \rbGTRS/ $\gtrs_\cm$ such that there is a run
\[
    \ctrconfig{\cmcontrol_0}{\ta_0}{\valszero}
    \tran{\empsym}
    \cdots
    \tran{\empsym}
    \ctrconfig{\cmcontrol_f}{\ta}{\vals}
\]
for some $\ta$ and $\vals$ iff there is a run
\[
    \cmconfig{\control_0}{0}{0}
    \cmtran \cdots \cmtran
    \cmconfig{\control_f}{0}{0}
\]
of $\cm$.
Since this latter problem is well-known to be undecidable, we obtain undecidability of control-state reachability for senescent \rbGTRS/.

In the following definition we use the following $1$-reversal-bounded counters:
$\ctrinc{0}, \ctrinc{1}, \ctrdec{0}$ and $\ctrdec{1}$.
We use
    $\rulesfresh$ to keep leaf nodes within the lifespan,
    $\rulesinc, \rulesdec$, and $\ruleszero$ to simulate the counter operations, and
    $\rulesfin$ to check
        $\ctrinc{i} = \ctrdec{i}$
        for both $i$ at the end of the run.
Furthermore, let
\[
    \begin{array}{rcl}
        \ap{\dvalsinc{i}}{\ctr}
        &=&
        \begin{cases}
            1 & \ctr = \ctrinc{i}
            \\
            0 & \text{otherwise,}
        \end{cases}
        \\
        \ap{\dvalsdec{i}}{\ctr}
        &=&
        \begin{cases}
            1 & \ctr = \ctrdec{i}
            \\
            0 & \text{otherwise, and}
        \end{cases}
        \\
        \ap{\dvalseq{i}}{\ctr}
        &=&
        \begin{cases}
            -1 & \ctr \in \set{\ctrinc{i},\ctrdec{i}}
            \\
            0 & \text{otherwise.}
        \end{cases}
    \end{array}
\]
Recall $\valszero$ maps all counters to zero.

Given a node $\tnode$ and trees $\tree_1, \ldots, \tree_n$, we will often
write $\tap{\tnode}{\tree_1, \ldots, \tree_n}$ to denote the tree with root
node $\tnode$ and left-to-right child sub-trees $\tree_1, \ldots, \tree_n$.
When $\tnode$ is labelled $\cha$, we may also write $\tap{\cha}{\tree_1,
\ldots, \tree_n}$ to denote the same tree.  We will often simply write $\cha$
to denote the tree with a single node labelled $\cha$.

For a tree $\tree$, let $\tasingle{\tree}$ be an NTA accepting only $\tree$.
For example, $\tasingle{\tap{\cha}{\chb}}$ is the automaton accepting only the
tree $\tap{\cha}{\chb}$, and $\tasingle{\cha}$ accepts only the tree containing
a single node labelled $\cha$.  Note, we do not use natural numbers as tree
labels, hence $\ta_1, \ta_2, \ldots$ may range over all NTAs.

\begin{nameddefinition}{def:cm-gtrs}{$\gtrs_\cm$}
    Given a two-counter machine
    $\cm = \tup{\cmcontrols, \cmrules}$
    and two control-states
    $\cmcontrol_0, \cmcontrol_f \in \cmcontrols$,
    we define a senescent \rbGTRS/ with lifespan $1$
    \[
        \gtrs_\cm = \tup{\controls, \alphabet, \oalphabet, \rules, \ctrs}
    \]
    where
    \[
        \begin{array}{rcl}
            \controls
            &=&
            \cmcontrols
            \uplus
            \setcomp{\cmfreeze{\cmcontrol}{i}}
                    {\cmcontrol \in \cmcontrols
                     \land
                     i \in \set{0,1}}
            \uplus
            \set{\fincontrol, \eqcontrol}
            \\
            \alphabet
            &=&
            \set{\nodeint, \nodegen, \nodenull, \nodectr{1}, \nodectr{2}}
            \\
            \oalphabet
            &=&
            \set{\empsym}
            \\
            \ctrs
            &=&
            \set{\ctrinc{1}, \ctrinc{2},
                 \ctrdec{1}, \ctrdec{2}}
            \\

            \rules
            &=&
            \rulesfresh \cup
            \rulesinc \cup
            \rulesdec \cup
            \ruleszero \cup
            \rulesfin
        \end{array}
    \]
    where
    \[
        \begin{array}{rcl}
            \rulesfresh
            &=&
            \setcomp{\ctrrule{\cmcontrol}{\tasingle{\tnode}}{\contrue}
                             {\empsym}
                             {\cmcontrol}{\tasingle{\tnode}}{\valszero}}
                    {\cmcontrol \in \cmcontrols
                     \land
                     \tnode \in \set{\nodegen, \nodectr{1}, \nodectr{2}}}
            \ \cup \\
            & &
            \setcomp{\ctrrule{\cmfreeze{\cmcontrol}{i}}
                             {\tasingle{\tnode}}
                             {\contrue}
                             {\empsym}
                             {\cmfreeze{\cmcontrol}{i}}
                             {\tasingle{\tnode}}
                             {\valszero}}
                    {\cmcontrol \in \cmcontrols
                     \land
                     \tnode
                     \in
                     \set{\nodegen, \nodectr{1}, \nodectr{2}}
                     \setminus
                     \set{\nodectr{i}}}
            \\
            \rulesinc
            &=&
            \setcomp{\ctrrule{\cmcontrol_1}
                             {\tasingle{\nodegen}}
                             {\contrue}
                             {\empsym}
                             {\cmcontrol_2}
                             {\tasingle{\tap{\nodeint}{\nodectr{i}, \nodegen}}}
                             {\dvalsinc{i}}}
                    {\cmrule{\control_1}{\cminc{i}}{\control_2} \in \cmrules}
            \\
            \rulesdec
            &=&
            \setcomp{\ctrrule{\cmcontrol_1}
                             {\tasingle{\nodectr{i}}}
                             {\contrue}
                             {\empsym}
                             {\cmcontrol_2}
                             {\tasingle{\nodenull}}
                             {\dvalsdec{i}}}
                    {\cmrule{\control_1}{\cmdec{i}}{\control_2} \in \cmrules}
            \\
            \ruleszero
            &=&
            \setcomp{
                \begin{array}{c}
                    \ctrrule{\cmcontrol_1}
                            {\tasingle{\nodegen}}
                            {\contrue}
                            {\empsym}
                            {\cmfreeze{\cmcontrol_2}{i}}
                            {\tasingle{\nodegen}}
                            {\valszero},
                    \\
                    \ctrrule{\cmfreeze{\cmcontrol_2}{i}}
                            {\tasingle{\nodegen}}
                            {\contrue}
                            {\empsym}
                            {\cmcontrol_2}
                            {\tasingle{\nodegen}}
                            {\valszero},

                \end{array}
            }{
                \cmrule{\control_1}{\cmzero{i}}{\control_2} \in \cmrules
            }
            \\
            \rulesfin
            &=&
            \set{\begin{array}{c}
                \ctrrule{\cmcontrol_f}
                        {\tasingle{\nodegen}}
                        {\contrue}
                        {\empsym}
                        {\eqcontrol}
                        {\tasingle{\nodegen}}
                        {\valszero},
                \\
                \ctrrule{\eqcontrol}
                        {\tasingle{\nodegen}}
                        {\contrue}
                        {\empsym}
                        {\eqcontrol}
                        {\tasingle{\nodegen}}
                        {\dvalseq{0}},
                \\
                \ctrrule{\eqcontrol}
                        {\tasingle{\nodegen}}
                        {\contrue}
                        {\empsym}
                        {\eqcontrol}
                        {\tasingle{\nodegen}}
                        {\dvalseq{1}},
                \\
                \ctrrule{\eqcontrol}
                        {\tasingle{\nodegen}}
                        {\ctrinc{0} = 0
                         \land
                         \ctrdec{0} = 0
                         \land
                         \ctrinc{1} = 0
                         \land
                         \ctrdec{1} = 0}
                        {\empsym}
                        {\fincontrol}
                        {\tasingle{\nodegen}}
                        {\valszero}
            \end{array}}
        \end{array}
    \]
\end{nameddefinition}
\OMIT{
\al{I was confused what $\valszero$ meant. Was it defined somewhere?}
\mh{Yep, but added a recap with the other defs}
}

\begin{namedproperty}{prop:cm-sim}{Simulation of $\cm$}
    For a given two-counter machine $\cm$ and control-states $\cmcontrol_0$ and $\cmcontrol_f$ there is a run
    \[
        \cmconfig{\control_0}{0}{0}
        \cmtran \cdots \cmtran
        \cmconfig{\control_f}{0}{0}
    \]
    of $\cm$ iff there is a run
    \[
        \ctrconfig{\cmcontrol_0}{\ta_0}{\valszero}
        \tran{\empsym}
        \cdots
        \tran{\empsym}
        \ctrconfig{\cmcontrol_f}{\ta}{\vals}
    \]
    for some $\ta$ and $\vals$ of $\gtrs_\cm$.
\end{namedproperty}
\begin{proof}
    Let
    $\cmcontrol_1 = \cmcontrol_0$
    and
    $\cmcontrol_n = \cmcontrol_f$ and suppose we have a run
    \[
        \cmconfig{\cmcontrol_1}{0}{0}
        \cmtran \cdots \cmtran
        \cmconfig{\cmcontrol_n}{0}{0} \ .
    \]
    We build the required run of $\gtrs_\cm$ by induction such that for configuration
    $\cmconfig{\cmcontrol_j}{\val^j_0}{\val^j_1}$
    along the run of $\cm$, we have a run to a configuration
    $\ctrconfig{\cmcontrol_j}{\ta_j}{\vals_j}$
    of $\gtrs_\cm$ such that
    \begin{itemize}
    \item
        there is one leaf node labelled $\nodegen$, this node has age $0$,
    \item
        the number of nodes $\nodectr{i}$ in $\ta_j$ is $\val^j_i$ for each
        $j \in \set{0, 1}$, each having age $0$, and
    \item
        $\ap{\vals_j}{\ctrinc{i}} -
         \ap{\vals_j}{\ctrinc{i}} =
         \val^j_i$
        for each
        $i \in \set{0,1}$.
    \end{itemize}
    In the base case the result holds trivially for the configuration
    $\ctrconfig{\cmcontrol_1}{\nodegen}{\valszero}$.
    Now take a transition
    $\cmconfig{\cmcontrol_j}{\cmop}{\cmcontrol_{j+1}}$
    from the run of $\cm$.
    By induction we have a run to
    $\ctrconfig{\cmcontrol_j}{\ta_j}{\vals_j}$
    as above.
    We show how to extend this run to
    $\ctrconfig{\cmcontrol_{j+1}}{\ta_{j+1}}{\vals_{j+1}}$.
    There are several cases depending on $\cmop$.
    In each case we show how to reach a tree satisfying the induction hypothesis, except the age of the leaf nodes.
    After the case analysis we show how to satisfy the age requirement also.
    \begin{itemize}
    \item
        When
        $\cmop = \cminc{i}$,
        we use
        $\ctrrule{\cmcontrol_j}
                 {\tasingle{\nodegen}}
                 {\contrue}
                 {\empsym}
                 {\cmcontrol_{j+1}}
                 {\tasingle{\tap{\nodeint}{\nodectr{i}, \nodegen}}}
                 {\dvalsinc{i}}$.
        It is easy to verify we reach
        $\ctrconfig{\cmcontrol_{j+1}}{\ta_{j+1}}{\vals_{j+1}}$
        as required.
    \item
        When
        $\cmop = \cmdec{i}$,
        we know the $i$th counter must have a value greater than zero, hence we can apply
        $\ctrrule{\cmcontrol_j}
                 {\tasingle{\nodectr{i}}}
                 {\contrue}
                 {\empsym}
                 {\cmcontrol_{j+1}}
                 {\tasingle{\nodenull}}
                 {\dvalsdec{i}}$.
        It is easy to verify we reach
        $\ctrconfig{\cmcontrol_{j+1}}{\ta_{j+1}}{\vals_{j+1}}$
        as required.
    \item
        When
        $\cmop = \cmzero{i}$,
        we know the $i$th counter must have value zero, hence there are no leaves labelled $\nodectr{i}$ in $\ta_j$.
        We can apply the following sequence of rules.
        \begin{enumerate}
        \item
            $\ctrrule{\cmcontrol_j}
                     {\tasingle{\nodegen}}
                     {\contrue}
                     {\empsym}
                     {\cmfreeze{\cmcontrol_{j+1}}{i}}
                     {\tasingle{\nodegen}}
                     {\valszero}$,
        \item
            $\ctrrule{\cmfreeze{\cmcontrol_{j+1}}{i}}
                     {\tasingle{\tnode}}
                     {\contrue}
                     {\empsym}
                     {\cmfreeze{\cmcontrol_{j+1}}{i}}
                     {\tasingle{\tnode}}
                     {\valszero}$
            to each leaf labelled by some
            $\tnode \in \set{\nodegen, \nodectr{0}, \nodectr{1}}
                        \setminus
                        \set{\nodectr{i}}$,
        \item
            $\ctrrule{\cmfreeze{\cmcontrol_{j+1}}{i}}
                     {\tasingle{\nodegen}}
                     {\contrue}
                     {\empsym}
                     {\cmcontrol_{j+1}}
                     {\tasingle{\nodegen}}
                     {\valszero}$.
        \end{enumerate}
        It is easy to verify we reach
        $\ctrconfig{\cmcontrol_{j+1}}{\ta_{j+1}}{\vals_{j+1}}$
        as required.
    \end{itemize}
    Finally, to obtain the age restriction on all leaf nodes, we apply
    $\ctrrule{\cmcontrol_{j+1}}
             {\tasingle{\tnode}}
             {\contrue}
             {\empsym}
             {\cmcontrol_{j+1}}
             {\tasingle{\tnode}}
             {\valszero}$
    to each leaf labelled by some
    $\tnode \in \set{\nodegen, \nodectr{0}, \nodectr{1}}$.

    Thus, by induction, we can reach a configuration
    $\ctrconfig{\cmcontrol_f}{\ta}{\vals}$
    such that, for each $i$ we have
    $\ap{\vals}{\ctrinc{i}} = \ap{\vals}{\ctrdec{i}}$.
    Thus, we can apply a sequence of rules from $\rulesfin$
    to reach
    $\ctrconfig{\fincontrol}{\ta}{\vals}$.
    In particular, we apply
    $\ctrrule{\cmcontrol_f}
             {\tasingle{\nodegen}}
             {\contrue}
             {\empsym}
             {\eqcontrol}
             {\tasingle{\nodegen}}
             {\valszero}$
    and then simultaneously reduce each reversal-bounded counter to zero using
    $\ctrrule{\eqcontrol}
             {\tasingle{\nodegen}}
             {\contrue}
             {\empsym}
             {\eqcontrol}
             {\tasingle{\nodegen}}
             {\dvalseq{i}}$
    repeatedly for each $i$, and then finally apply
    \[
        \ctrrule{\eqcontrol}
                {\tasingle{\nodegen}}
                {\ctrinc{0} = 0
                 \land
                 \ctrdec{0} = 0
                 \land
                 \ctrinc{1} = 0
                 \land
                 \ctrdec{1} = 0}
                {\empsym}
                {\fincontrol}
                {\tasingle{\nodegen}}
                {\valszero}
    \]
    to complete this direction of the proof.

    We prove the opposite direction via two inductions.
    First, take some run of $\gtrs_\cm$, which necessarily has the form
    \[
        \ctrconfig{\control_1}{\ta_1}{\vals_1}
        \tran{\empsym}
        \cdots
        \tran{\empsym}
        \ctrconfig{\control_n}{\ta_n}{\vals_n}
        \tran{\empsym}
        \ctrconfig{\eqcontrol}{\ta_n}{\vals_n}
        \tran{\empsym}
        \cdots
        \tran{\empsym}
        \ctrconfig{\eqcontrol}{0}{0}
        \tran{\empsym}
        \ctrconfig{\fincontrol}{0}{0}
    \]
    where
        the last sequence of transitions (from $\control_n$) are all from $\rulesfin$,
        $\control_1 = \cmcontrol_0$,
        $\ta_1 = \nodegen$,
        $\vals_1 = \valszero$, and
        $\control_n = \cmcontrol_f$.
    Let
    $\numleaves{\ta}{i}$
    be the number of leaves labelled $\nodectr{i}$ in $\ta$.
    We first prove by induction over the run that for all
    $1 \leq j \leq n$
    and
    $i \in \set{0,1}$
    we have
    $\numleaves{\ta_j}{i} = \ap{\vals_j}{\ctrinc{i}} - \ap{\vals_j}{\ctrdec{i}}$.
    This is a straightforward induction that can be seen by observing
    \begin{itemize}
    \item
        the base case is immediate,
    \item
        all rules in
        $\rulesfresh \cup \ruleszero$ do not change
        $\numleaves{\ta_j}{i}$,
        $\ap{\vals_j}{\ctrinc{i}}$, or
        $\ap{\vals_j}{\ctrdec{i}}$,
    \item
        all rules in $\rulesinc$ increase both
        $\numleaves{\ta_j}{i}$, and
        $\ap{\vals_j}{\ctrinc{i}}$,
        by one, and leave
        $\ap{\vals_j}{\ctrdec{i}}$
        unchanged,
    \item
        all rules in $\rulesdec$ decrease
        $\numleaves{\ta_j}{i}$
        by one, increase
        $\ap{\vals_j}{\ctrdec{i}}$
        by one, and leave
        $\ap{\vals_j}{\ctrinc{i}}$,
        unchanged, and
    \item
        there are no rules from $\rulesfin$.
    \end{itemize}
    Given
    $\numleaves{\ta_j}{i} = \ap{\vals_j}{\ctrinc{i}} - \ap{\vals_j}{\ctrdec{i}}$
    for all $j$ and $i$, we construct, also by induction, a sequence
    \[
        \cmconfig{\cmcontrol_1}{\val^1_0}{\val^1_1},
        \ldots,
        \cmconfig{\cmcontrol_n}{\val^n_0}{\val^n_1}
    \]
    of $\cm$ such that for all $j$ and $i$ we have
    $\numleaves{\ta_j}{i} = \val^j_0$
    and
    $\control_j \in \set{\cmcontrol_j,
                         \cmfreeze{\cmcontrol_j}{0},
                         \cmfreeze{\cmcontrol_j}{1}}$
    and, either
    \begin{itemize}
    \item
        $\cmconfig{\cmcontrol_j}{\val^j_0}{\val^j_1}
         \cmtran
         \cmconfig{\cmcontrol_{j+1}}{\val^{j+1}_0}{\val^{j+1}_1}$
        is a transition of $\cm$, or
    \item
        $\cmconfig{\cmcontrol_j}{\val^j_0}{\val^j_1}
         =
         \cmconfig{\cmcontrol_{j+1}}{\val^{j+1}_0}{\val^{j+1}_1}$.
    \end{itemize}
    In the base case we set
    $\cmconfig{\cmcontrol_1}{\val^1_0}{\val^1_0}
     =
     \cmconfig{\cmcontrol_0}{0}{0}$.
    Next, take a transition
    \[
        \ctrconfig{\control_j}{\ta_j}{\vals_j}
        \tran{\empsym}
        \ctrconfig{\control_{j+1}}{\ta_{j+1}}{\vals_{j+1}}
    \]
    of $\gtrs_\cm$.
    There are several cases depending on which rule $\rruler$ was applied.
    \begin{itemize}
    \item
        If
        $\rruler \in \rulesfresh$
        then we set
        $\cmconfig{\cmcontrol_j}{\val^j_0}{\val^j_1}
         =
         \cmconfig{\cmcontrol_{j+1}}{\val^{j+1}_0}{\val^{j+1}_1}$
        and the properties follow from
        $\cmconfig{\cmcontrol_j}{\val^j_0}{\val^j_1}$
        by induction.

    \item
        If
        $\rruler \in \rulesinc$
        then for some $i$ we have
        $\rruler = \ctrrule{\cmcontrol_j}
                           {\tasingle{\nodegen}}
                           {\contrue}
                           {\empsym}
                           {\cmcontrol_{j+1}}
                           {\tasingle{\tap{\nodeint}{\nodectr{i}, \nodegen}}}
                           {\dvalsinc{i}}$
        and
        $\cmrule{\cmcontrol_j}{\cminc{i}}{\cmcontrol_{j+1}}$
        is a rule of $\cm$.
        We apply this rule to obtain
        $\cmconfig{\cmcontrol_j}{\val^j_0}{\val^j_1}
         \cmtran
         \cmconfig{\cmcontrol_{j+1}}{\val^{j+1}_0}{\val^{j+1}_1}$
        and we can directly verify
        $\numleaves{\ta_{j+1}}{i} = \val^{j+1}_i$
        for each $i$ as required.

    \item
        If
        $\rruler \in \rulesdec$
        then for some $i$ we have
        $\rruler = \ctrrule{\cmcontrol_j}
                           {\tasingle{\nodectr{i}}}
                           {\contrue}
                           {\empsym}
                           {\cmcontrol_{j+1}}
                           {\tasingle{\nodenull}}
                           {\dvalsdec{i}}$
        and
        $\cmrule{\cmcontrol_j}{\cmdec{i}}{\cmcontrol_{j+1}}$
        is a rule of $\cm$.
        We apply this rule to obtain
        $\cmconfig{\cmcontrol_j}{\val^j_0}{\val^j_1}
         \cmtran
         \cmconfig{\cmcontrol_{j+1}}{\val^{j+1}_0}{\val^{j+1}_1}$
        and we can directly verify
        $\numleaves{\ta_{j+1}}{i} = \val^{j+1}_i$
        for each $i$ as required.

    \item
        If
        $\rruler \in \ruleszero$
        there are two sub-cases.
        \begin{itemize}
        \item
            In the first case, for some $i$ we have
            $\rruler = \ctrrule{\cmcontrol_j}
                               {\tasingle{\nodegen}}
                               {\contrue}
                               {\empsym}
                               {\cmfreeze{\cmcontrol_{j+1}}{i}}
                               {\tasingle{\nodegen}}
                               {\valszero}$
            and
            $\cmrule{\cmcontrol_j}{\cmzero{i}}{\cmcontrol_{j+1}}$
            is a rule of $\cm$.
            If we apply this rule we obtain
            $\cmconfig{\cmcontrol_j}{\val^j_0}{\val^j_1}
             \cmtran
             \cmconfig{\cmcontrol_{j+1}}{\val^{j+1}_0}{\val^{j+1}_1}$
            and we can directly verify
            $\numleaves{\ta_{j+1}}{i} = \val^{j+1}_i$
            for each $i$ as required.
            However, we need to prove
            $\cmrule{\cmcontrol_j}{\cmzero{i}}{\cmcontrol_{j+1}}$
            can be applied.
            That is, we need to prove $\val^j_i$ is zero.
            Here we use
            $\numleaves{\ta_{j'}}{i} =
             \ap{\vals_{j'}}{\ctrinc{i}} - \ap{\vals_{j'}}{\ctrdec{i}}$
            for all $j'$.
            From the definition of $\gtrs_\cm$ we know that the run from
            $\ctrconfig{\cmfreeze{\cmcontrol_{j+1}}{i}}{\ta_{j+1}}{\vals_{j+1}}$
            must eventually reach $\cmcontrol_{j+1}$ where
            $\cmfreeze{\cmcontrol_{j+1}}{i}$
            is the only control-state seen before $\cmcontrol_{j+1}$ is reached.
            During this time, we cannot refresh any node labelled $\nodectr{i}$.
            Thus, assume for contradiction that $\val^j_i$ is not zero.
            Since
            $\numleaves{\ta_{j}}{i} = \val^j_i$
            we know there is at least one leaf labelled $\nodectr{i}$.
            Since this node cannot refresh while the control-state is
            $\cmfreeze{\cmcontrol_{j+1}}{i}$
            this node will have age $2$ once $\cmcontrol_{j+1}$ is reached.
            Thus, since the lifespan is $1$, this node cannot be rewritten by the end of the run.
            This means $\ta_n$ has at least one node labelled $\nodectr{i}$.
            Since
            $1 \leq \numleaves{\ta_n}{i} =
             \ap{\vals_n}{\ctrinc{i}} - \ap{\vals_n}{\ctrdec{i}}$
            we know
            $\ap{\vals_n}{\ctrinc{i}} \neq \ap{\vals_n}{\ctrdec{i}}$.
            However, the final transitions of the run of $\gtrs_\cm$ use rules from $\rulesfin$ and have the effect of ensuring
            $\ap{\vals_n}{\ctrinc{i}} = \ap{\vals_n}{\ctrdec{i}}$.
            Hence, we have a contradiction, and $\val^j_i = 0$.
            Thus we can apply
            $\cmrule{\cmcontrol_j}{\cmzero{i}}{\cmcontrol_{j+1}}$
            as needed.

        \item
            If
            $\rruler = \ctrrule{\cmfreeze{\cmcontrol_j}{i}}
                               {\tasingle{\nodegen}}
                               {\contrue}
                               {\empsym}
                               {\cmcontrol_{j+1}}
                               {\tasingle{\nodegen}}
                               {\valszero}$
            we set
            $\cmconfig{\cmcontrol_j}{\val^j_0}{\val^j_1}
             =
             \cmconfig{\cmcontrol_{j+1}}{\val^{j+1}_0}{\val^{j+1}_1}$
            which satisfies the required properties since
            $\cmconfig{\cmcontrol_j}{\val^j_0}{\val^j_1}$
            did by induction.
        \end{itemize}
    \end{itemize}
    Thus, we have a sequence
    $\cmconfig{\cmcontrol_1}{\val^1_0}{\val^1_1},
     \ldots,
     \cmconfig{\cmcontrol_n}{\val^n_0}{\val^n_1}$
    from which we can immediately extract a run of $\cm$
    from
    $\cmconfig{\cmcontrol_1}{\val^1_0}{\val^1_1} = \cmconfig{\cmcontrol_0}{0}{0}$
    to
    $\cmconfig{\cmcontrol_n}{\val^n_0}{\val^n_1}
     =
     \cmconfig{\cmcontrol_f}{\val^n_0}{\val^n_1}$.
    That
    $\val^n_0 = \val^n_1 = 0$
    follows since the final transitions from $\cmcontrol_n$ have the effect of asserting
    $\ap{\vals_n}{\ctrinc{i}} - \ap{\vals_n}{\ctrdec{i}} = 0$
    from which we conclude
    $\numleaves{\ta_n}{i} = 0$
    and since
    $\val^n_i = \numleaves{\ta_n}{i}$
    we complete the proof as required.
\end{proof}

Thus, via \refproperty{prop:cm-sim} we can reduce the reachability problem for two-counter machines to the control-state reachability problem for senescent \rbGTRS/.
Thus, we show the control-state reachability problem is undecidable and complete the proof of Theorem~\ref{thm:senescent-undec}.